\documentclass[a4paper,fleqn]{cas-dc}
\usepackage[numbers,sort&compress]{natbib}
\usepackage{gensymb}
\pdfoutput=1
\let\printorcid\relax

\def\tsc#1{\csdef{#1}{\textsc{\lowercase{#1}}\xspace}}
\tsc{WGM}
\tsc{QE}

\begin{document}
\let\WriteBookmarks\relax
\def\floatpagepagefraction{1}
\def\textpagefraction{.001}

\shorttitle{Large-scale atomistic simulation of dislocation core structure in face-centered cubic metal with Deep Potential method}   
\shortauthors{Deng}  
\title [mode = title]{Large-scale atomistic simulation of dislocation core structure in face-centered cubic metal with Deep Potential method}  

\author[1,2]{Fenglin Deng}[orcid=0000-0002-6443-9173]
\fnmark[1]
\author[1,2]{Hongyu Wu}[orcid=0000-0002-7696-934X]
\fnmark[1]
\author[1,2]{Ri He}
\author[3]{Peijun Yang}
\author[1,2,4]{Zhicheng Zhong}[orcid=0000-0003-1507-4814]
\cormark[1]
\fntext[1]{These authors contribute equally to this work.}
\cortext[1]{Corresponding author: zhong@nimte.ac.cn}


\affiliation[1]{organization={CAS Key Laboratory of Magnetic Materials and Devices},
            addressline={Ningbo Institute of Materials Technology and Engineering, Chinese Academy of Sciences}, 
            city={Ningbo},
            postcode={315201}, 
            country={China}}

\affiliation[2]{organization={Zhejiang Province Key Laboratory of Magnetic Materials and Application Technology},
            addressline={Ningbo Institute of Materials Technology and Engineering, Chinese Academy of Sciences}, 
            city={Ningbo},
            postcode={315201}, 
            country={China}}

\affiliation[4]{organization={China Center of Materials Science and Optoelectronics Engineering},
            addressline={University of Chinese Academy of Sciences}, 
            city={Beijing},
            postcode={100049}, 
            country={China}}

\affiliation[3]{organization={Northeastern University},
            addressline={School of Materials Science and Engineering}, 
            city={Shenyang},
            postcode={110819}, 
            country={China}}

\nonumnote{}

\begin{abstract}
The core structure of dislocations is critical to their mobility, cross slip, and other plastic behaviors. Atomistic simulation of the core structure is limited by the size of first-principles density functional theory (DFT) calculation and the accuracy of classical molecular dynamics with empirical interatomic potentials. Here, we utilize a Deep Potential (DP) method learned from DFT calculations to investigate the dislocations of face-centered cubic copper on a large scale and obtain their core structures and energies. The validity of the DP description of the core structure and elastic strain from dislocation is confirmed by a fully discrete Peierls model. Moreover, the DP method can be further extended easily to dislocations with defects such as surface or vacancy, and our study will pave a way in the large-scale atomistic simulation of dislocation on the DFT level.
\end{abstract}


\maketitle

\section{Introduction}\label{intro}
Dislocation is one of the most important defects which determines plastic properties in metals \cite{hirth1982theory}. Dislocation usually has a detailed core structure related to crystal properties such as elasticity and atomic bonding \cite{vitek1974,duesbery1991,vitek2004}. From Peierls' dislocation theory \cite{peierls1940}, the formation of core structure originates from a balance in which the elastic interaction between dislocation density tends to make the core wider, while the misfit energy is the complete opposite tendency. For example, compared with covalent crystals, breaking metallic bonds will induce small misfit energy which means a wider dislocation core \cite{hirth1982theory}. In face-centered cubic (FCC) metal, determination of dislocation width is a key issue because it has a large impact on the cross slip process \cite{escaig1968cross,bonneville1988study,rasmussen1997}. Investigating the dislocation core at atomic level will serve to indicate the experimental signatures of core effects. 

For investigating the dislocation core structure, Peierls-Nabarro (P-N) model together with the $\gamma$-surface is a successful analytical model \cite{vitek1974,peierls1940,nabarro1947,vitek1968,bulatov1997}. But this model cannot predict the atomic positions around the dislocation core precisely and cannot handle anisotropic material \cite{schoeck20057} or complex systems such as high entropy alloy in a satisfactory way. For atomic simulation, first-principles density functional theory (DFT) calculations can predict the atomic structures with quantum accuracy. Nevertheless, DFT calculation is limited by small length and time scale. The splitting width of the dissociated dislocation in FCC copper is up to $2\sim4$ $\mathrm{nm}$ from experimental observations \cite{stobbs1971, weiler1995high}, which is beyond the reaches of conventional DFT methods \cite{rodney2017}. Interatomic potentials are a very effective method for understanding the core structure but predicted dislocation core usually depends on the quality of potential \cite{groger2009,chiesa2009}. In addition, developing an interatomic potential for the alloy system with acceptable accuracy is not a trivial task. Accurate atomic simulations of dislocation core structure demand a generalizable potential with high accuracy.

In recent years, machine learning methods have been used as a powerful tool to develop the interatomic potential of crystalline materials \cite{behler2007ge, bartok2010, zhao2020snap, schutt2018schnet,wang2018dpkit}. Among them, the recently proposed Deep Potential (DP) method based on a deep neural network (DNN) can provide a DFT-level accurate interatomic potential \cite{zhang2018dp,zhang2021pd}. Many high accurate DP potentials have been developed for systems of vastly different materials \cite{zhang2018end,fu2021deep,he2022structural} including metals and alloys \cite{zhang2020,jiang2021}. Zhang \textit{et al.} developed a DP potential of Cu, and the accuracy of this potential has been validated. This potential outperforms the modified embedded atom method (MEAM) potentials in almost all examined properties including elasticity and stacking fault energy \cite{zhang2020}. We try to use it in studying the dislocation core structure in copper.

\begin{figure}[t]
 \centering
    \includegraphics[width=8.35cm,height=4.64cm]{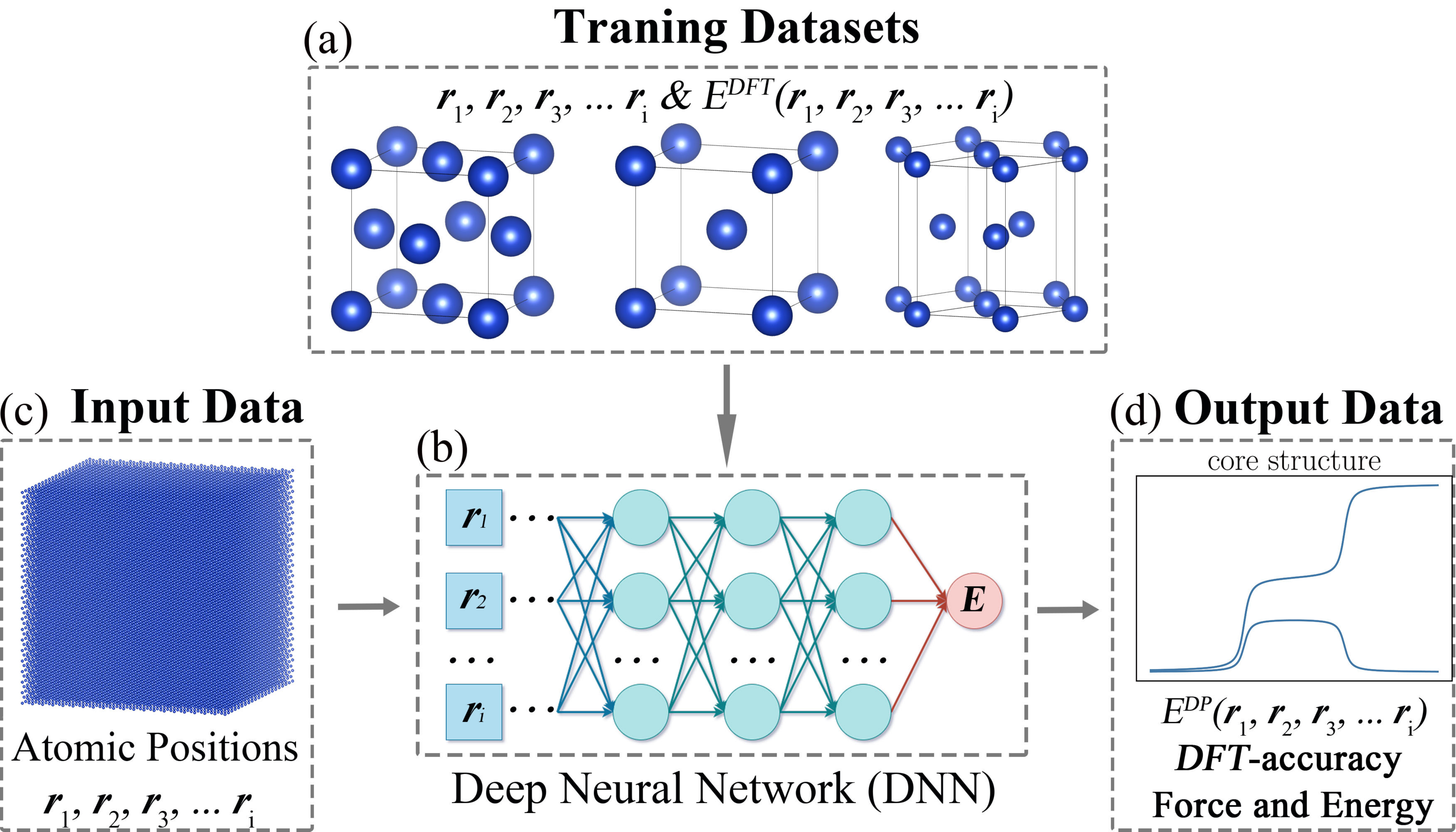}
   \caption{(a) A large dataset of DFT energies and forces of a wide range of atomic configurations. (b) The deep neural network trained by Deep Potential (c) and (d) The neural network potential can be used to predict the energy and force of large supercell that contains the vacancy, surface, and dislocation at DFT level.}\label{dp}
\end{figure}

{In the DP method, the neural network is trained from a large dataset that contains a wide range of atomic configurations with a small number of atoms (Figure~\ref{dp}(a)). Each lattice configuration is labeled by atomic coordinates and the corresponding DFT energy and atomic forces. The energy calculation of variable size supercells from DNN is implemented by setting up a local environment for every atom and its neighbors inside a cutoff radius. The DP method will not output the total energy of configuration directly but return atomic energies determined by an atom's local environment.} By summing atomic energies, the well-trained DP potential can be used to predict the energy and force of a large defected supercell at DFT level (Figure~\ref{dp}(b)).

In this paper, we use DP method to investigate the $1/2{\langle 110 \rangle} \{111\}$ dislocations in FCC copper. The core structure and energy of dislocation are calculated by the DP method. Since it is almost impossible to validate the accuracy of dislocation properties obtained from the DP method by comparing DFT results, we implement the calculation of extended dislocation core structure and energy by the fully discrete Peierls model (see section \ref{csdd}) developed by Wang \textit{et al.} \cite{wang2015,wang2016,xiang2020}. The core structure predicted by the DP method agrees well with which obtained from the discrete model. This result demonstrates the generalizability of the DP potential of Cu. Furthermore, by analyzing the energies of dislocation arrays with different sizes, the DP method is proven to reproduce the elastic interaction between dislocations on a large scale. We also investigate the properties of a screw dislocation in Cu film and vacancy-dislocation interaction in the bulk. DP method provides significant promise for studying dislocations at the atomic level and also offers critical physical quantities for other simulation methods on a larger scale, such as discrete dislocation dynamics and phase-field simulation \cite{lesar2014,beyerlein2016under,bertin2020}.

\section{Method}
\subsection{Construction of supercell}\label{scl}

\begin{figure}
 \centering
    \includegraphics[width=8.5cm,height=5.66cm]{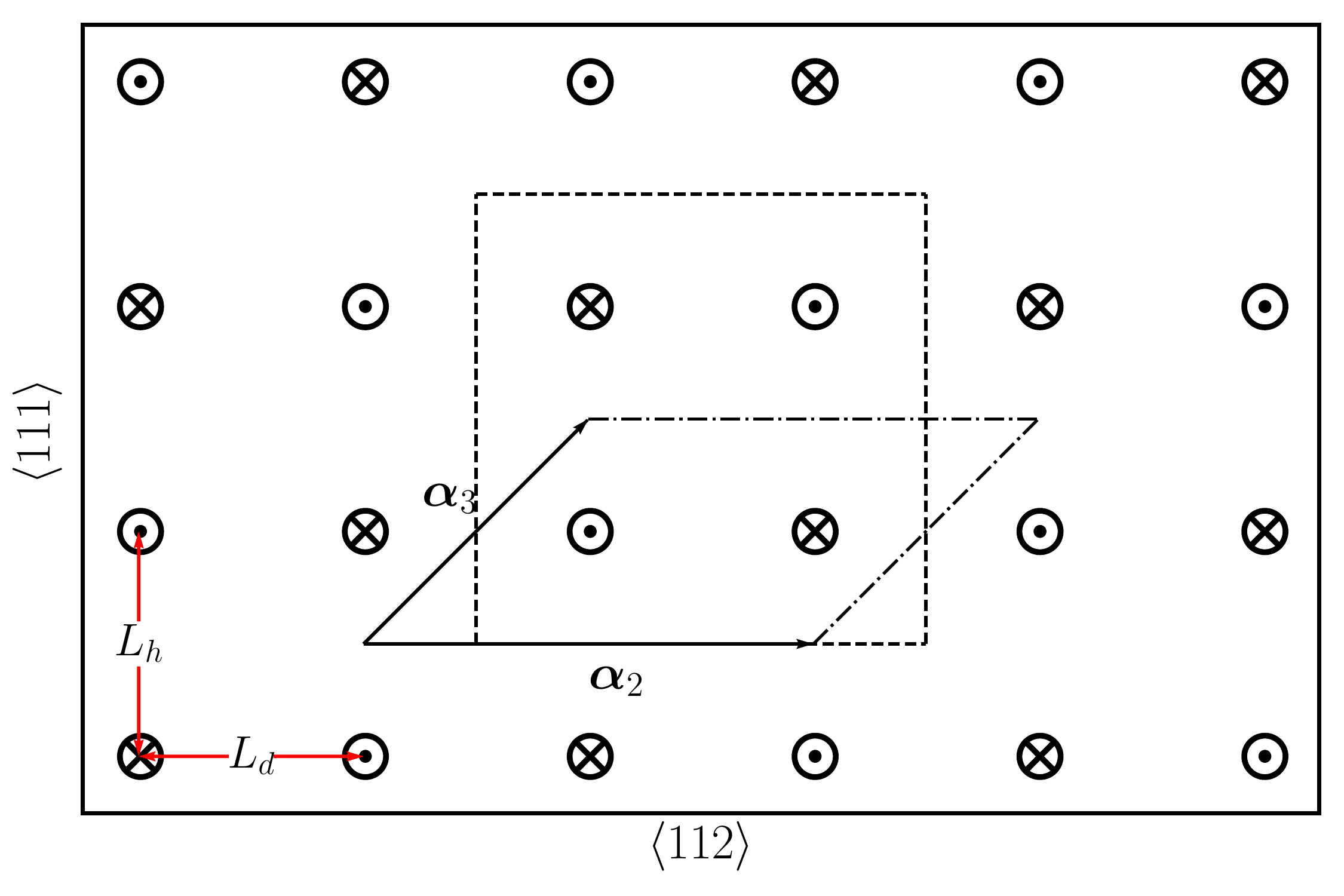}
   \caption{Schematic diagram of the screw dislocation quadrupole array. The dashed line shows a quadrupole supercell and the dash-dotted one shows a dipole supercell.}\label{sc}
\end{figure}

We investigate the $1/2{\langle 110 \rangle}\{111\}$ dislocation which possesses $\{111\}$ glide plane and $1/2{\langle 110 \rangle}$ Burgers vector. By choosing different dislocation line directions, the $0\degree$ (screw), $60\degree$ (mixed), $90\degree$ (edge), and $30\degree$ (mixed) straight dislocation is constructed. The periodic boundary conditions (PBCs) are used in atomic simulations. Figure~\ref{sc} shows two different simulation cells in this work. Symbols $\odot$ and $\otimes$ represent straight dislocations with opposite Burgers vectors. The dashed line shows a supercell that contains a dislocation quadrupole with two glide planes, while the dash-dotted one contains a dislocation dipole with only one glide plane. They both form a dislocation quadrupole array in the plane perpendicular to the dislocation line. The initial dislocations are created by the displacement fields proposed in \cite{deng2019}, which is the exact solution of dislocation and anti-dislocation array in the P-N model.

To describe the simulated supercells explicitly, the following auxiliary lattice vectors $\mathbf{a}_i$ are used,
\begin{align}
\mathbf{a}_1&=\frac{1}{2}{\left[011\right]}, & \mathbf{a}_2&=\frac{1}{2}{[2\overline{1}1]},& \mathbf{a}_3&={[11\overline{1}]}.\nonumber
\end{align}
The vectors $\mathbf{a}_1$ and $\mathbf{a}_2$ represent two unequivalent choices of dislocation line direction (see Figure~\ref{discrew}(a)). The vector $\mathbf{a}_3$ is the normal direction of glide plane. The basis vectors of quadrupole supercells can be defined by vectors $\boldsymbol{\alpha}_i=l_i\mathbf{a}_i$. In the case of dipole supercell, the basis vector $\boldsymbol{\alpha}_1$ and $\boldsymbol{\alpha}_2$ are the same with quadrupole case, while $\boldsymbol{\alpha}_3$ depends on the direction of dislocation line. If ${\langle 110 \rangle}$ is chosen as the dislocation line direction, we set $\boldsymbol{\alpha}_3=l_3\mathbf{a}_3+(l_2\mathbf{a}_2+\mathbf{b})/2$, where $\mathbf{b}$ is the Burgers vector. For the ${\langle 112 \rangle}$ case, we set $\boldsymbol{\alpha}_3=l_3\mathbf{a}_3+(l_1\mathbf{a}_1+\mathbf{b})/2$. Now we can use the parameters array $(l_1, l_2, l_3)$ to represent the supercell in calculation.

For testing the generalizability of the Cu DP potential, the quadrupole supercells with $(l_1, l_2, l_3)=(1, 8, 4)$ for $0\degree$ and $60\degree$ dislocation, and $(8, 1, 4)$ for $90\degree$ and $30\degree$ dislocation are constructed. Each supercell has 192 atoms. In addition, the supercell $(l_1, l_2, l_3)=(2, 1, 2)$ with two free surfaces or a single vacancy is also taken into account. 

To investigate the dislocation core structure, we construct the dislocation dipole supercells with $(l_1, l_2, l_3)=(1, 480, 70)$ for screw and $60\degree$ dislocation, and $(480, 1, 70)$ for edge dislocation. These cells which contain about $10^5$ atoms are large enough to obtain stable dislocation cores.

\subsection{Computation details}

To obtain the lattice configurations in testing dataset, we performed molecular dynamics calculations by Vienna \textit{ab initio} Simulation Package (VASP) code\cite{kresse199615,kresse199610}. The Perdew-Burke-Ernzerhof functional \cite{PBEsol2009} was adopted for structure relaxation. The kinetic energy cutoff was set to 650 eV, the K-point was set using the Monkhorst–Pack mesh \cite{monkhorst1976} with the spacing 0.1 $\mathrm{\mathring{A}}^{-1}$, and the temperature was set to 300 K. We also performed static calculations by VASP to obtain the generalized stacking fault energy.

We performed the molecular statics (MS) calculation by LAMMPS code \cite{plimpton19951} to obtain the stable structure of dislocation cores. The DP potential of Cu comes from the recent work by Zhang \textit{et al.}\cite{zhang2020}, and EAM potential is proposed by Mishin \textit{et al.} \cite{mishin2001}. We have attempted to update the original Cu DP potential by adding generalized stacking fault configurations to the training dataset for obtaining a specified potential in studying dislocations. This modification is not evident, so the original Cu DP potential is used in this work (see section~\ref{gamma} for details). 

\section{Results}

\subsection{Accuracy of DP method}
\begin{figure}[h]
 \centering
    \includegraphics[width=8.5cm,height=8.5cm]{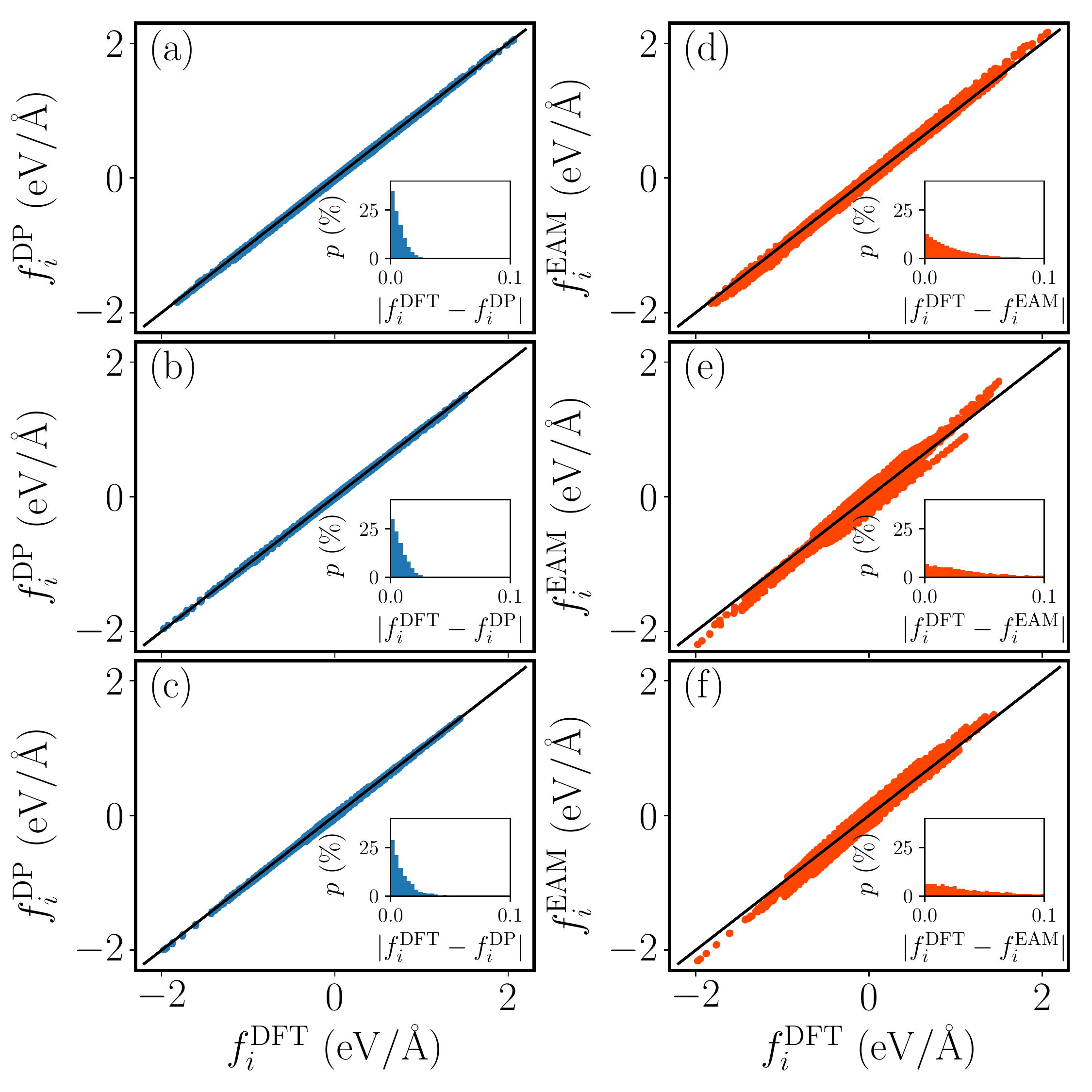}
   \caption{(a), (b), (c) ((d), (e), (f)) show the atomic force obtained from DP (EAM potential) compared with DFT in systems contain the dislocations, free surfaces, the Cu vacancy respectively. The subplots show the relative frequency $p$ of $|f_{i}^{\mathrm{DP/EAM}}-f_{i}^{\mathrm{DFT}}|$ for each case.}\label{force}
\end{figure}

The investigation of atomic simulation of dislocation demands the atomic potential possessing high accuracy in calculating atomic force and generalized stacking fault energy. In \cite{zhang2020}, only elastic properties of Cu bulk and formation energies of vacancy, surface, and intrinsic stacking fault (ISF) obtained from the DP method are compared with those obtained from other methods. For demonstrating the accuracy of DP potential, we compare the atomic forces calculated by DP and EAM potential with DFT results. The atomic forces in different configurations are shown in Figure~\ref{force}. Considering the limitations of DFT method, the configurations containing dislocations are all small and unstable. The atomic forces predicted by DP agree well with DFT results, which are much better than the EAM results. The root-mean-squared errors (RMSEs) of atomic force in free surface systems obtained from the DP method with respect to DFT references are one order of magnitude smaller than EAM results (see Table~\ref{tb1}). For the systems containing dislocations or Cu vacancy, the RMSEs of DP results are about four times smaller than EAM results. The comparison of total energies shows a similar trend as well. Especially for the vacancy case, the RMSEs of energies from the DP method are about forty times smaller than EAM results. 
\begin{table}[h!]
\begin{center}
\caption{The root-mean-squared errors (RMSEs) of atomic force from DP and EAM with respect to DFT references in dislocation, free surface, and vacancy systems.}\label{tb1}
\begin{tabular}{l|c|c|c} 
\toprule
RMSEs $(\mathrm{eV/\mathring{A}})$& dislocation & free surface & vacancy \\
\midrule
DP  & $8.81\times 10^{-3}$ & $9.75\times 10^{-3}$ & $1.28\times 10^{-2}$ \\
EAM & $3.23\times 10^{-2}$ & $7.19\times 10^{-2}$ & $6.2 \times 10^{-2}$ \\
\bottomrule
\end{tabular}
\end{center}
\end{table}

\subsection{Generalized stacking fault energy}\label{gamma}

The generalized stacking fault energy (GSFE) or $\gamma$-surface is extremely useful in qualitatively analyzing the spreading of a dislocation core, which is defined by the surplus energy per unit area when a relative gliding exists between two half infinite bulks\cite{vitek1968,vitek1974}. The negative gradient of GSFE describes the restoring stress between the mismatched lattice planes, which is necessary for the P-N model.

For interatomic potentials, the ability to calculate an accurate GSFE is rather essential because the predicted dislocation core is associated with the character of obtained GSFE \cite{DUESBERY19981481}. Therefore, we test the accuracy of the original Cu DP potential for predicting GSFE in this section. We have attempted to update the original DP potential by including generalized stacking faults in the training dataset. The GSFE calculated by original and updated DP potential and EAM potential is compared with that obtained from DFT.

For the $\{111\}$ glide plane in FCC copper, the GSFE along two directions, ${\left[011\right]}$ and ${[2\overline{1}1]}$, is taken into consideration. The normal direction of the glide plane, ${[11\overline{1}]}$, is neglected. The structure relaxation is allowed along the direction perpendicular to the glide plane before energy calculation. We use the slip displacement field $\mathbf{s}$ to represent the rigid glide vector of generalized stacking fault which is defined by the relative displacement field of the two mismatched lattice planes, $\mathbf{s}=\mathbf{u}^{\mathrm{a}}-\mathbf{u}^{\mathrm{b}}$. The $\mathbf{u}^{\mathrm{a}}$ $(\mathbf{u}^{\mathrm{b}})$ denotes the displacement field of the atom which belongs to the lattice plane above (below) the glide plane. The component of $\mathbf{s}$ along ${\left[011\right]}$ is denoted by $s_{x}$ and ${[2\overline{1}1]}$ by $s_{y}$. Due to the high cost of DFT calculation, only several positions along these two glide directions are calculated.  

\begin{figure}[h]
 \centering
    \includegraphics[width=8.5cm,height=12.32cm]{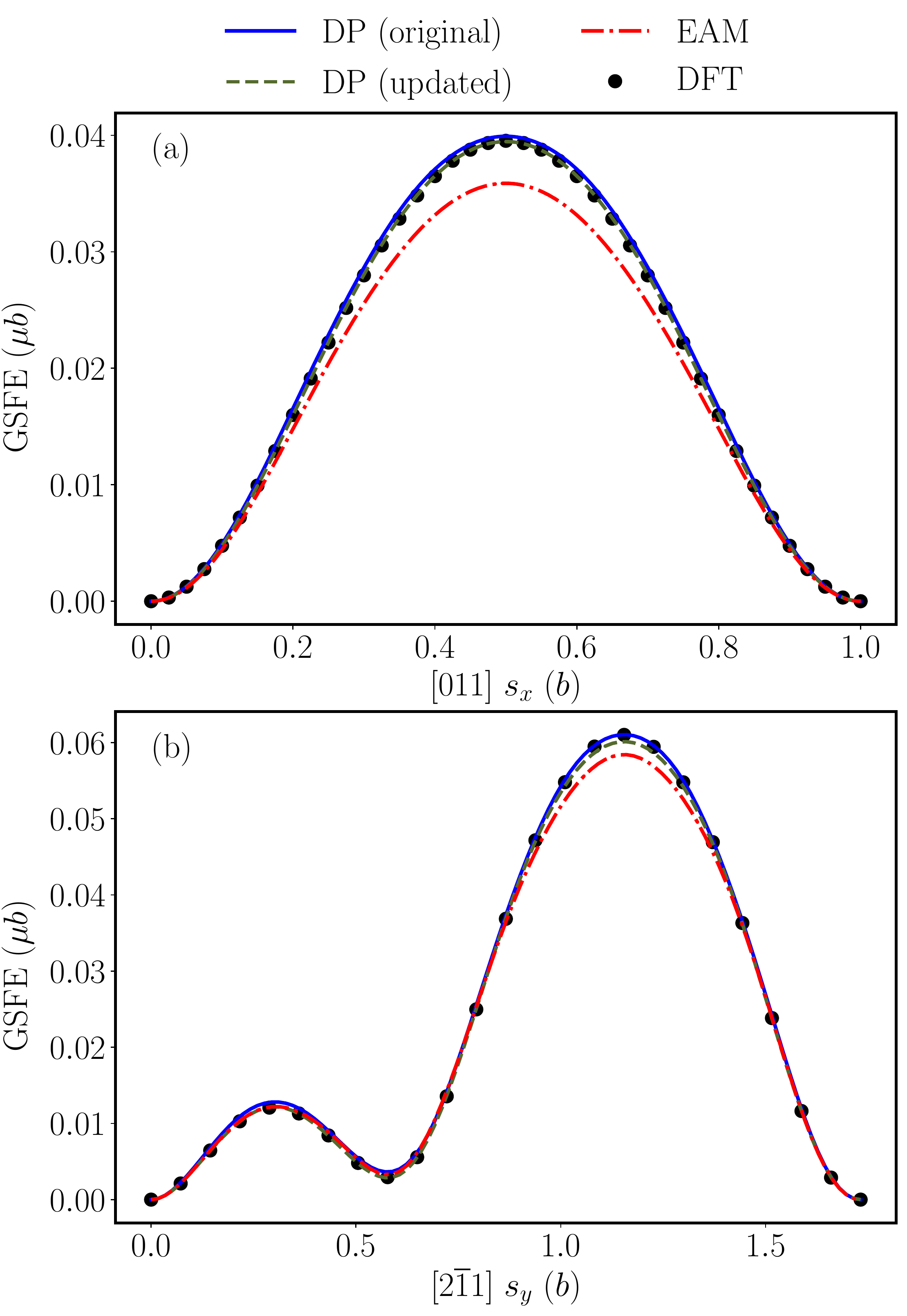}
   \caption{The generalized stacking fault energy (GSFE) along (a) ${\left[011\right]}$ and (b) ${[2\overline{1}1]}$ direction is calculated by DFT, original and updated DP, and EAM potential.}\label{gamma-surface}
\end{figure}

Figure~\ref{gamma-surface} shows the GSFE calculated by DFT, original and updated DP, and EAM potential. There is no evident difference between original and updated DP potential in predicting GSFE. Although the training dataset contains no generalized stacking fault, the GSFE predicted by the original DP potential is almost the same as that by DFT. When the rigid gliding happens around the initial or the ISF state ($\mathbf{s}=0$ or $s_y=b/\sqrt{3}$, where $b=|\mathbf{b}|=2.57 \mathrm{\mathring{A}}$), both DP and EAM potential predict the same GSFE as DFT does. When the two half bulk glide around the most unstable positions, like $s_x=b/2$ or $s_y=2b/\sqrt{3}$ shown in Fig \ref{gamma-surface}, the GSFE predicted by DP seems more accurate than that from EAM potential. Due to the original DP potential performing so well on the calculation of GSFE, we believe it can be applied to simulate the extended dislocations in FCC copper directly.

\subsection{Core structure of dissociated dislocations}\label{csdd}

\begin{figure}[h]
 \centering
    \includegraphics[width=8.5cm,height=7.484cm]{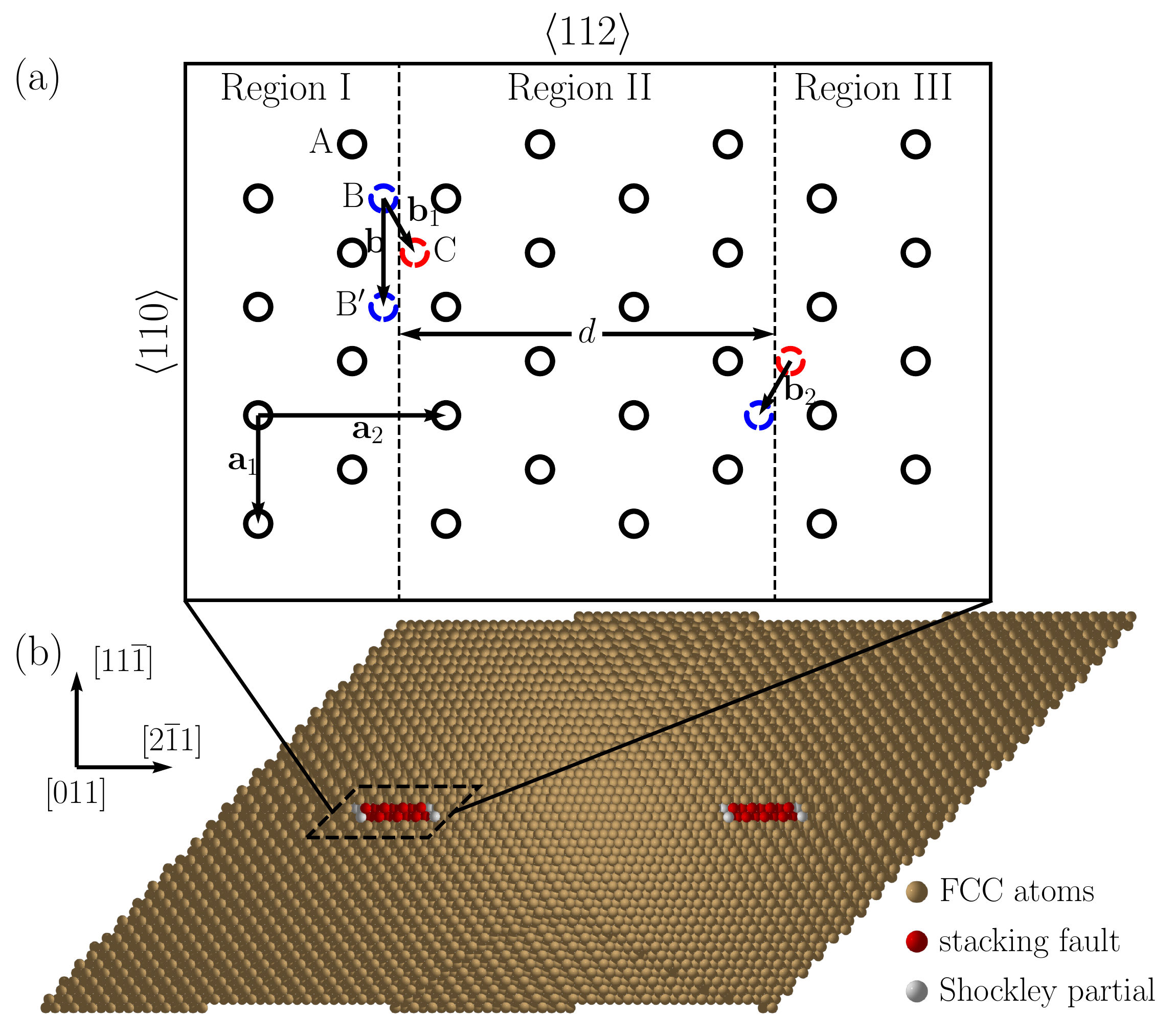}
   \caption{(a) Schematic diagram of the dissociation mechanism of the screw dislocation in FCC lattice. The dashed lines represent the dislocation line of two Shockley partial dislocations. (b) A $(l_1,l_2,l_3)=(2,40,15)$ supercell contains 7200 Cu atoms and two dissociated screw dislocations.}\label{discrew}
\end{figure}

Figure~\ref{discrew}(a) shows the well-known dissociation mechanism of a full screw dislocation with Burgers vector $\mathbf{b}$. The mismatching exists between the $\{111\}$ lattice planes with atom positions labeled as A (black circle) and B (blue dashed circle) respectively. A full screw dislocation can be constructed by gliding the atoms on B-plane from B to B$^{\prime}$ relatively when region changes from $\mathrm{\uppercase\expandafter{\romannumeral1}}$ to $\mathrm{\uppercase\expandafter{\romannumeral3}}$. Due to the relatively small unstable stacking fault (USF) energy in $\gamma$-surface of copper, the atoms at B prefer gliding to position C (red dashed circle) firstly and then from C to B$^{\prime}$. Which means the $1/2{\langle 110 \rangle}$ dislocation in copper would dissociate to two Shockley $1/6{\langle 112 \rangle}$ partial dislocations which are separated by an ISF ribbon (labeled as region $\mathrm{\uppercase\expandafter{\romannumeral2}}$ in Figure~\ref{discrew}(a)). The distance between two Shockley partial dislocations is called the splitting width (denoted by $d$).

\begin{figure*}
 \centering
    \includegraphics[width=17cm,height=12.714cm]{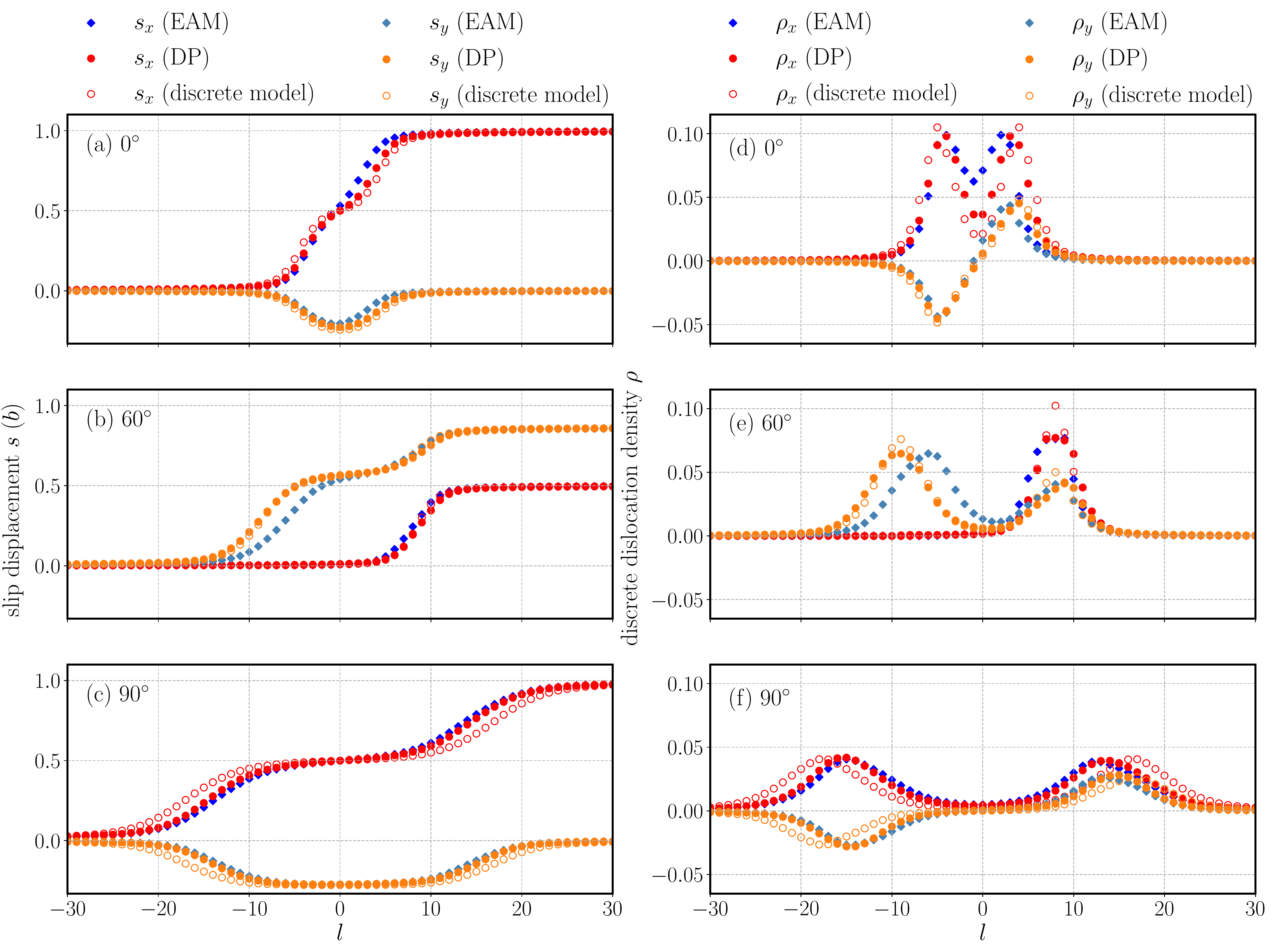}
   \caption{The slip displacement and discrete density of $0\degree$, $60\degree$, and $90\degree$ dislocation are obtained from EAM potential, DP method, and fully discrete Peierls model which are plotted in solid and empty circles respectively.}\label{struc}
\end{figure*}

In this section, we investigate the core structure of $1/2{\langle 110 \rangle}\{111\}$ dislocations by DP, EAM potential, and the fully discrete Peierls model.  The $0\degree$, $60\degree$, and $90\degree$ dislocation are investigated and the $30\degree$ dislocations are not considered because they tend to annihilate under PBCs. To obtain the stable dislocation core by DP or EAM potential, a large distance between dislocations in the quadrupole array is set and MS calculations are performed. As shown in Figure~\ref{discrew}(b), the $0\degree$ screw dislocations in supercell have split into Shockley partials automatically. The atoms around dislocation lines are plotted in white and stacking fault ribbons are plotted in red.

Main information of dislocation core is contained in the slip displacement $\mathbf{s}(l)$, where integer $l$ labels the atoms on the mismatched lattice plane. Due to the planar character of these extended dislocations, only the components belong to glide plane are taken into account. We define the discrete dislocation density $\rho_{i}(l)$ by slip displacement field, $\rho_{i}(l)=s_{i}(l+1)-s_{i}(l)$. The splitting width $d$ is defined by the distance between two peaks of dislocation density ($\rho_{x}$ in screw and edge case and $\rho_{y}$ in mixed case). The slip displacement $s_{i}(l)$ ($i=x,y$) extracted from the DP and EAM results is shown in Figure~\ref{struc}.

For evaluating core structures obtained from MS, we use the fully discrete Peierls model to study the slip displacements. DFT methods are not considered because of the large splitting width of dislocation cores in Cu. When dislocation line is along ${\langle 110 \rangle}$ direction, the energy functional in this discrete model is 
\begin{align}
F=&\frac{1}{4\lambda_{x}^2}\sum_{l=-\infty}^{\infty}\left[\beta_{s}\rho_{x}^2(l)+\beta_{e}\rho_{y}^2(l) \right] \nonumber \\ 
&-\frac{1}{4\pi\lambda_{x}}\sum_{l,l^{\prime}=-\infty}^{\infty}\left[K_{s}\rho_{x}(l)\rho_{x}(l^{\prime})+K_{e}\rho_{y}(l)\rho_{y}(l^{\prime}) \right]\times \nonumber \\ 
&\psi^{(0)}\left(|l-l^{\prime}|+\frac{1}{2}\right)+\sum_{l=-\infty}^{\infty}\gamma(s_x,s_y), \label{ef}
\end{align}
where $\lambda_x$ is the step length defined by the distance between the lattice lines paralleled to ${\langle 110 \rangle}$, $K_s$ and $K_e$ are the energy pre-factors of screw and edge dislocations, $\beta_s$ and $\beta_e$ are the contact-interaction constants, $\psi^{(0)}(x)$ is the first derivative of the logarithm of the gamma function, and $\gamma(s_x,x_y)$ is the $\gamma$-surface. When ${\langle 112 \rangle}$ is chosen as the dislocation line direction, the energy functional can be obtained by exchanging the dislocation density $\rho_x$ and $\rho_y$ and changing the $\lambda_x$ to $\lambda_y$ in (\ref{ef}), where $\lambda_y$ denotes the step length between the lattice lines paralleled to ${\langle 112 \rangle}$. For the $\{111\}$ glide plane in FCC lattice, the step length $\lambda_x=\sqrt{3}b/2$ and $\lambda_y=b/2$. The energy pre-factors $K_s=\mu$ and $K_e=\mu/(1-\nu)$ \cite{hirth1982theory}, where $\mu$ is the shear modulus and $\nu$ is the Poisson ratio. The contact-interaction constants are determined by the following formulae \cite{wang2008dis},
\begin{align*}
\beta_s&=\frac{3}{4}\left(1-\tan^2\theta\sin^2\phi\right)\mu h,\\
\beta_e&=\frac{3}{4}\left(\frac{2-2\nu}{1-2\nu}-\tan^2\theta\cos^2\phi\right)\mu h,
\end{align*}
where $\tan\theta=1/\sqrt{2}$, $\phi=\pi/6$, and $h=\sqrt{2/3}b$ is the distance of the two nearest lattice planes paralleled to glide plane.

To compute the shear modulus $\mu$ and Poisson ratio $\nu$ under the isotropic approximation, the elastic constant calculated by DFT in \cite{zhang2020} is used. The fitting formula for $\gamma(s_x,s_y)$ in \cite{vorontsov2012} is applied because an analytical formula for $\gamma$-surface in the dislocation equations is necessary. By applying the numerical method mentioned in \cite{karlin2000numerical}, slip displacements determined by energy functional (\ref{ef}) are obtained and shown in Figure~\ref{struc}. 


We find these different methods predict similar core structures for respective dislocations (see Figure \ref{struc}). In Table \ref{tbs}, the splitting widths obtained from these methods are compared with {experimental values}. The EAM potential tends to predict a narrower splitting width in the $0\degree$ case and the discrete model tends to predict a wider core in the $90\degree$ case. The splitting width $d$ (in unit of $b$) of $0\degree$, $60\degree$, and $90\degree$ dislocation calculated by DP method is 6.1, 14.7, and 15 respectively, which is close to the {experiment values} \cite{stobbs1971,weiler1995high} or results in recent works \cite{liu2017156,liu201769}. In general, these results show that the DP method can predict the dissociated dislocation core structure with high accuracy.

\begin{table}[h!]
\begin{center}
\caption{Comparison of splitting width $d$ (in units of $b$) obtained from different methods for dislocations in Cu.}\label{tbs}
\begin{tabular}{l|c|c|c|c} 
\toprule
    & EAM & DP & discrete model & experiment values\\
\midrule
$0\degree$  & 5.2 & 6.1 & 7.8 &7.0 \cite{stobbs1971} \\
$60\degree$ & 13.0 & 14.7 & 14.7 & 13.3 \cite{weiler1995high}\\
$90\degree$ & 14.5 & 15.0 &17.5  &14.9 \cite{stobbs1971}\\
\bottomrule
\end{tabular}
\end{center}
\end{table}

\subsection{Core energy of screw dislocation}

We study the core energy and elastic energy of a screw dislocation in FCC copper by DP method and fully discrete Peierls model in this section. From continuum theory, elastic energy caused by dislocations originates from the logarithmic form interaction. This elastic energy of the dislocation quadrupole array is divergent when increasing dislocation distances and system size simultaneously. The core energy can be obtained by subtracting the elastic part from the total energy of a screw dislocation.

\begin{figure}[t]
 \centering
    \includegraphics[width=8.5cm,height=11.33cm]{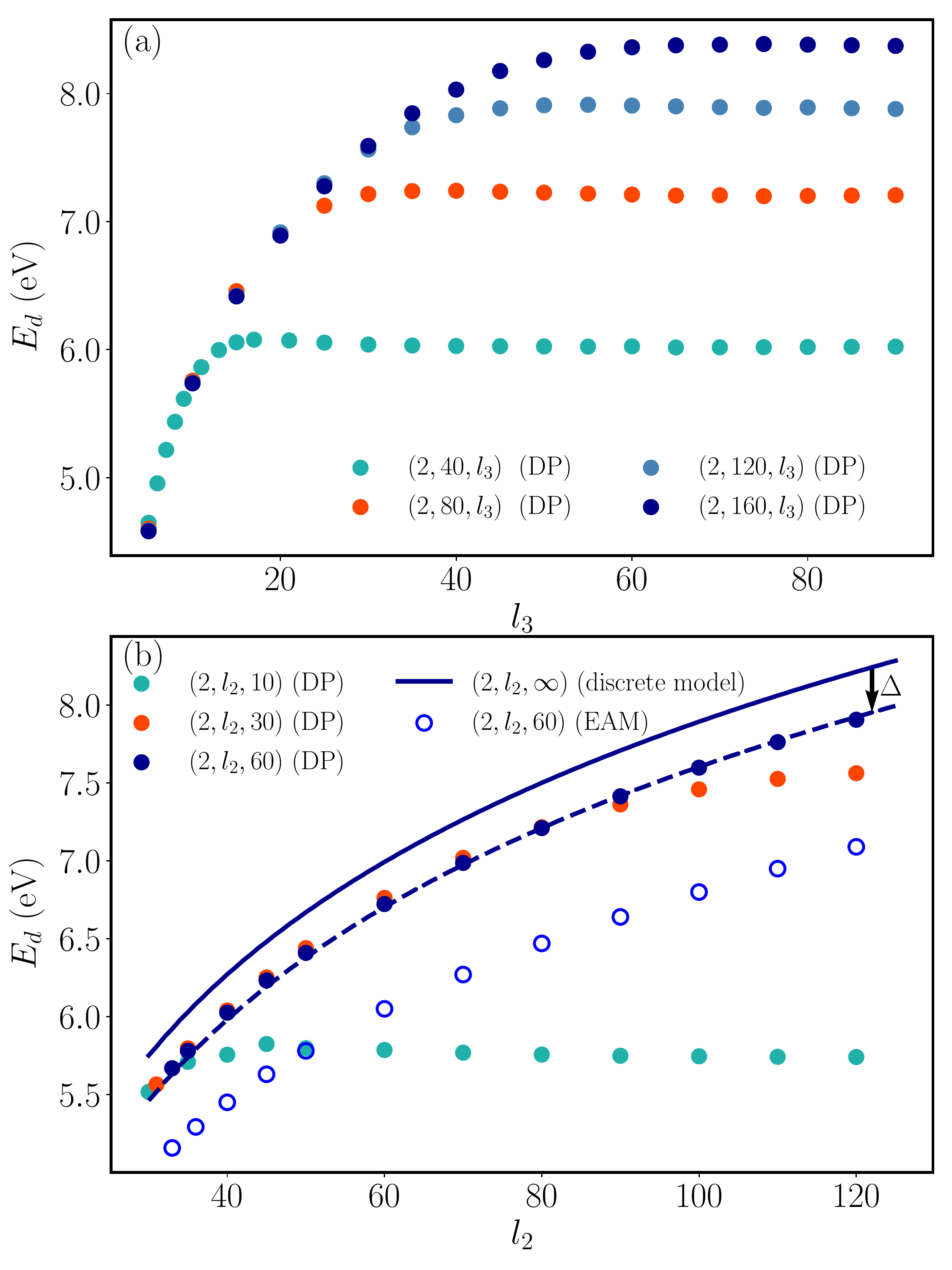}
   \caption{The energy of dislocation array varies with increasing distance in direction (a) ${[11\overline{1}]}$ or (b) ${[2\overline{1}1]}$. DP method can describe the logarithmic elastic interaction between dislocations. The dislocation energy obtained from the fully discrete Peierls model is plotted by the solid line (or dashed line with a constant shift).}\label{energy}
\end{figure}

In DP method, we can obtain the total dislocation energy of the $(l_1, l_2, l_3)$ supercell by subtracting energy $N_aE_a$ from that of defected supercell, where $N_a=6l_1l_2l_3$ is the number of Cu atoms in the defected system, $E_a=-3.728$ $\mathrm{eV}$ is the atomic energy in FCC copper from DP potential. As shown in Figure \ref{energy}(a), we increase the cell parameter $l_3$ from $5$ to $90$ with fixed $l_2=40$, $80$, $120$, or $160$. For a specified distance of dislocation dipole $L_d$, a large enough $L_h$ of supercell is necessary for ensuring the convergence of dipole energy. To investigate the distance $L_d$ dependence of dislocation elastic energy, it's enough to set $l_3=60$ when changing $l_2$ from $30$ to $120$. 

For analyzing the results obtained from DP method, we use fully discrete Peierls model to calculate the energy of a dipole array that is equivalent to a quadrupole array with infinite large $L_h$. Specifically, we consider the energy caused by one screw dislocation in the $(l_1,l_2,l_3\rightarrow \infty)$ array where all dislocations share a common glide plane. The elastic energy per unit length of one screw dislocation in it is
\begin{align*}
E_{elastic}=&-\frac{1}{4\pi}\sum_{l,l^{\prime}=1}^{l_2}\left[K_s\rho_x(l)\rho_x(l^{\prime})+K_e\rho_y(l)\rho_y(l^{\prime}) \right]\times\\
&\sum_{i=-\infty}^{\infty}(-1)^{i}\psi^{(0)}\left(|l-l^{\prime}+il_2|+\frac{1}{2}\right).
\end{align*}
where the infinity summation is truncated by a proper large number which ensures the convergence of results. Due to the locality of contact-interaction and misfit interaction, it's reasonable to define the core energy $E_{core}$ (unit length) as follows,
\begin{flalign*}
&E_{core}=\frac{1}{4\lambda_{x}}\sum_{l=1}^{l_2}\left[\beta_{s}\rho_{x}^2(l)+\beta_{e}\rho_{y}^2(l) \right]+\lambda_{x}\sum_{l=1}^{l_2}\gamma(s_x,s_y).&
\end{flalign*}
Then the total energy of a dislocation dipole array $(l_1, l_2, \infty)$ is 
\begin{align}
E_d=2l_1b(E_{elastic}+E_{core}). \label{dpe}
\end{align}
It's hard to solve slip displacement $\mathbf{s}(l)$ in dipole array self-consistently. Therefore we use the isolated dislocation result obtained in the previous section as an approximate solution.

The dislocation energy obtained from (\ref{dpe}) is also shown in Fig \ref{energy}(b). Compared with the discrete model result, the DP result shows a similar divergent tendency which demonstrates the long-range dislocation interaction holds in the DP method. The energy predicted by the discrete model is a little bit larger than that from the DP method. This energy deviation might come from the inaccurate core structure described by the discrete model. If this model result is moved down by about $\Delta=0.29$ eV (about $0.028$ $\mathrm{eV/\mathring{A}}$ for each screw dislocation), it will agree well with the energy from the DP method. With the assistance of the discrete model, we estimate the unit length core energy $E_{core}$ of a screw dislocation in copper is about $0.22$ $\mathrm{eV/\mathring{A}}$. These results indicate that DP can describe the long-range elastic interaction caused by dislocations and can provide valuable physical quantities such as core energy for other simulation methods.

\subsection{Surface effect and vacancy-dislocation interaction}
The above sections only concern ideal isolated dislocations or periodic dislocation array. For interpreting actual plastic properties, how other defects affect dislocation properties is very important. In this section, we study the interactions between dislocation and some intrinsic defects such as free surface and Cu vacancy. 

The free surface plays a vital role in nanoscale materials and it has been studied by atomic simulations or in the Peierls' framework \cite{dutta2008,cheng2012,bai2016}. The core size and mobility of dislocation usually vary with film thickness. The vacancy-dislocation interaction is another key issue in dislocation dynamics because it governs vacancy diffusion in the vicinity of dislocation and controls the dislocation climb. This interaction has been modeled by elastic theory and atomic simulations \cite{bullough1970,clouet20063543}. Yet investigating the surface effect or vacancy interaction of dislocations in copper by DFT is difficult because the defected system is too large. The analytical theory or classical interatomic potentials can not describe these defects precisely. Therefore DP is a proper method for studying the free surface and Cu vacancy and this method is used in this section.

\begin{figure}[t]
 \centering
    \includegraphics[width=8.5cm,height=9.8cm]{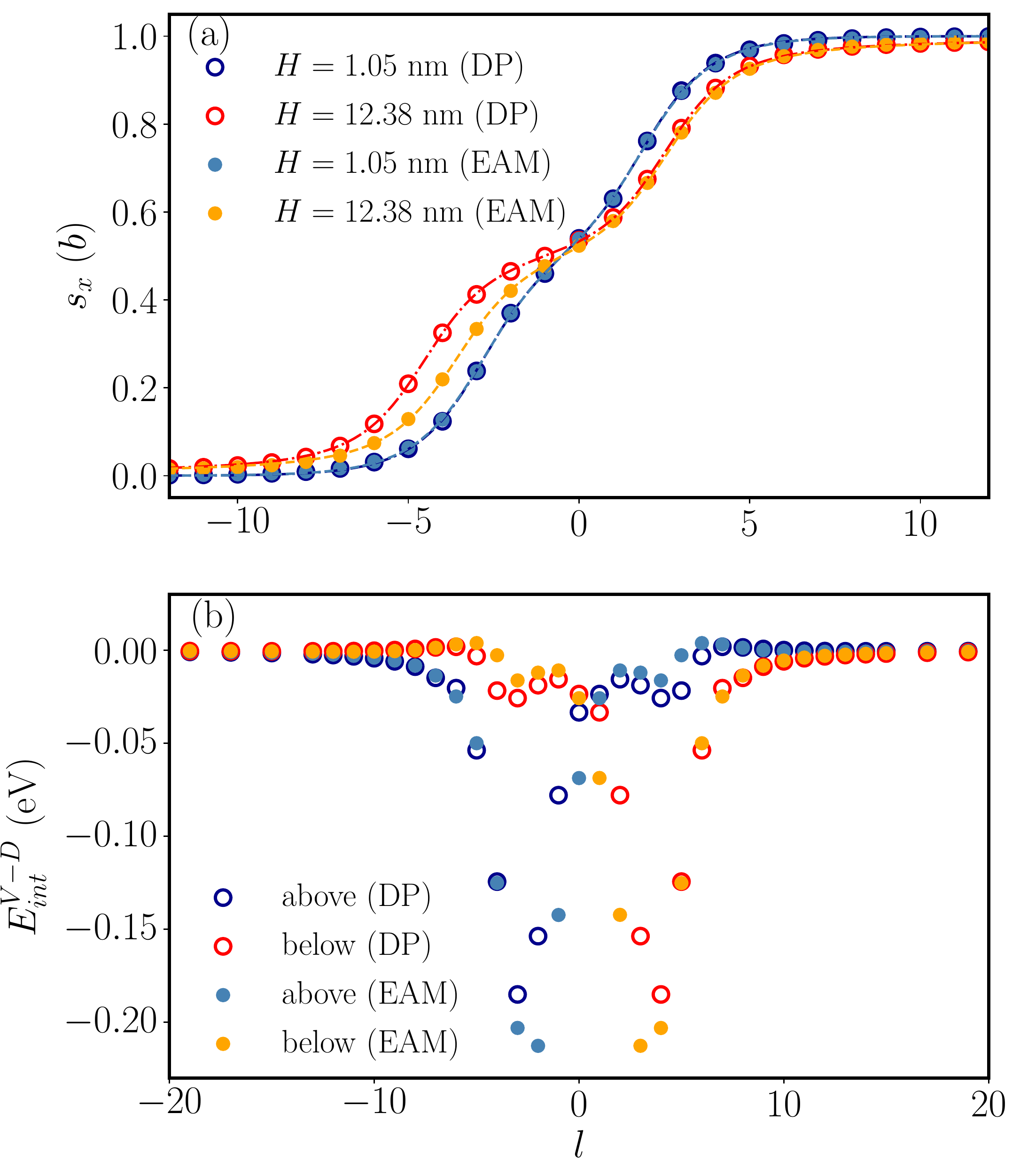}
   \caption{(a) The screw dislocation core structure in Cu films with different thicknesses. The thinner film is, the smaller splitting width is. (b) The interaction energy between the vacancy and the screw dislocation in FCC copper.}\label{film}
\end{figure}

Firstly, we investigate the surface effect of a screw dislocation in Cu films. The free surfaces of a Cu film are two paralleled $\{111\}$ lattice planes. The middle plane of a film is chosen as the dislocation glide plane. Four cases with different film thickness $H=1.05$, $1.47$, $2.31$, and $12.38$ $\mathrm{nm}$ are considered. Slip displacement fields $s_x$ are extracted from the stable configurations and shown in Figure~\ref{film}(a). The splitting widths $d$ and Peierls stresses $\tau_p$ are listed in Table~\ref{tb2}. The core structure of a screw dislocation in a free-standing Cu film becomes narrower than that in an FCC copper bulk when the film thickness is smaller than $2$ $\mathrm{nm}$ (about 10 $\{111\}$ lattice planes). For the film with thickness $H=12.38$ $\mathrm{nm}$, Peierls stress is about $7.5$ MPa which is close to the bulk case ($2.9$ MPa obtained by molecular dynamics in \cite{liu201769}). If the film thickness is smaller than 3 $\mathrm{nm}$, the Peierls stress of screw dislocation will be up to several hundred MPa which is much larger than that in bulk. In general, the surface will significantly affect the core and mobility of screw dislocation when the thickness of Cu film is about several nanometers.

The interaction between a single vacancy and a screw dislocation $E_{int}^{V-D}$ in copper is studied by the DP method as well. We calculate the energy of a dipole supercell $(l_1, l_2, l_3)=(4,160,70)$ containing a single vacancy in the lattice plane above or below the glide plane. The zero point of $E_{int}^{V-D}$ is set by the energy of the case that Cu vacancy is far from the dislocation. As shown in Figure~\ref{film}(b), the maximum attractive energy is -0.185 eV in the compressive region of the partial dislocation. In contrast to this strong attractive energy, the vacancy-dislocation interaction observed in the tensile region is much smaller.

\begin{table}[h!]
\begin{center}
\caption{Surface effect on the splitting width and Peierls stress of screw dislocations in different Cu films.}\label{tb2}
\begin{tabular}{l|c|c|c|c} 
\toprule
Film thickness $H$ ($\mathrm{nm}$) & 1.05 & 1.47 & 2.31 &12.38 \\
\midrule
$d$ ($b$)\qquad\qquad (DP) & 4.0 & 4.9 & 5.9 & 6.1 \\
$\tau_p$ (MPa) \qquad(DP) & 892.5 & 501.3 & 244.9 & 7.5 \\
\midrule
$d$ ($b$)\qquad\qquad (EAM) & 4.0 & 4.5 & 5.2 & 5.3 \\
$\tau_p$ (MPa) \qquad (EAM) & 916.6 & 589.7 & 210.2 & 5.2 \\
\bottomrule
\end{tabular}
\end{center}
\end{table}

\section{Conclusions}\label{conclu}
In summary, we use the DP method to investigate $1/2{\langle 110 \rangle}\{111\}$ dislocation in FCC copper. The DP method can predict the energies and atomic forces for large dislocation systems on the DFT level, which is more accurate than EAM results. The DP predictions of splitting width of $0\degree$, $60\degree$, and $90\degree$ dislocation is close to experiment values. With the assistance of the fully discrete Peierls model, we analyze the results of the screw dislocation array obtained by the DP method. The long-range elastic behavior of dislocation can be described by the DP method well. By subtracting the elastic part from total energy, we estimate the core energy of screw dislocation in copper is $0.22$ $\mathrm{eV/\mathring{A}}$. 

In addition, the DP method demonstrates its advantages in describing the effects of intrinsic defects on dislocation. We observed significant surface effects on screw dislocation when Cu film thickness is about several nanometers. When the film thickness is smaller than $3$ $\mathrm{nm}$, the Peierls stress of screw dislocation will be up to several hundred MPa. For the vacancy-dislocation interaction in copper, the maximum attractive energy between vacancy and compressive side of Shockley partial dislocation is -0.185 eV. We believe the DP method will open a new avenue in studying the kink, jog, or pinning in dislocation dynamics. Moreover, compared to empirical interatomic potentials, the deep learning based DP method is easier to extend and more accessible in complex systems including high entropy alloy.

\section{Acknowledgement}\label{akno}
We acknowledge financial support from the National Key R$\&$D Program of China (Grant No. 2021YFA0718900, and No. 2017YFA0303602), the Key Research Program of Frontier Sciences of CAS (Grant No. ZDBS-LY-SLH008), the National Nature Science Foundation of China (Grants No. 11974365), the Science Center of the National Science Foundation of China (52088101), and K.C. Wong Education Foundation (GJTD-2020-11). Calculations were performed at the Supercomputing Center of Ningbo Institute of Materials Technology and Engineering.

\section{Data availability}
The Cu DP potential used in this work is available in the online open data repository \url{http://dplibrary.deepmd.net/}.



\begin{thebibliography}{53}
\expandafter\ifx\csname natexlab\endcsname\relax\def\natexlab#1{#1}\fi
\providecommand{\url}[1]{\texttt{#1}}
\providecommand{\href}[2]{#2}
\providecommand{\path}[1]{#1}
\providecommand{\DOIprefix}{doi:}
\providecommand{\ArXivprefix}{arXiv:}
\providecommand{\URLprefix}{URL: }
\providecommand{\Pubmedprefix}{pmid:}
\providecommand{\doi}[1]{\href{http://dx.doi.org/#1}{\path{#1}}}
\providecommand{\Pubmed}[1]{\href{pmid:#1}{\path{#1}}}
\providecommand{\bibinfo}[2]{#2}
\ifx\xfnm\relax \def\xfnm[#1]{\unskip,\space#1}\fi
\bibitem[{Hirth and Lothe(1982)}]{hirth1982theory}
\bibinfo{author}{J.~P. Hirth}, \bibinfo{author}{J.~Lothe},
  \bibinfo{title}{Theory of dislocations}, \bibinfo{publisher}{New York:
  McGraw-Hill}, \bibinfo{year}{1982}.
\bibitem[{Vitek(1974)}]{vitek1974}
\bibinfo{author}{V.~Vitek},
\newblock \bibinfo{title}{Theory of the core structures of dislocations in
  body-centered-cubic metals},
\newblock \bibinfo{journal}{Crystal Lattice Defects} \bibinfo{volume}{5}
  (\bibinfo{year}{1974}) \bibinfo{pages}{1--34}.
\bibitem[{Duesbery and Richardson(1991)}]{duesbery1991}
\bibinfo{author}{M.~S. Duesbery}, \bibinfo{author}{G.~Y. Richardson},
\newblock \bibinfo{title}{The dislocation core in crystalline materials},
\newblock \bibinfo{journal}{Critical Reviews in Solid State and Materials
  Sciences} \bibinfo{volume}{17} (\bibinfo{year}{1991}) \bibinfo{pages}{1--46}.
\bibitem[{Vitek(2004)}]{vitek2004}
\bibinfo{author}{V.~Vitek},
\newblock \bibinfo{title}{Core structure of screw dislocations in body-centred
  cubic metals: relation to symmetry and interatomic bonding},
\newblock \bibinfo{journal}{Philosophical Magazine} \bibinfo{volume}{84}
  (\bibinfo{year}{2004}) \bibinfo{pages}{415--428}.
\bibitem[{Peierls(1940)}]{peierls1940}
\bibinfo{author}{R.~Peierls},
\newblock \bibinfo{title}{The size of a dislocation},
\newblock \bibinfo{journal}{Proceedings of the Physical Society}
  \bibinfo{volume}{52} (\bibinfo{year}{1940}) \bibinfo{pages}{34--37}.
\bibitem[{Escaig(1968)}]{escaig1968cross}
\bibinfo{author}{B.~Escaig},
\newblock \bibinfo{title}{Cross-slipping process in the f.c.c. structure},
\newblock in: \bibinfo{editor}{A.~B.~J. A.R.~Rosenfield, G.T.~Hahn},
  \bibinfo{editor}{R.~Jaffee} (Eds.), \bibinfo{booktitle}{Proceedings of the
  Battelle Colloquium in Dislocation Dynamics}, \bibinfo{publisher}{New York:
  McGraw-Hill}, \bibinfo{year}{1968}, pp. \bibinfo{pages}{655--677}.
\bibitem[{Bonneville et~al.(1988)Bonneville, Escaig, and
  Martin}]{bonneville1988study}
\bibinfo{author}{J.~Bonneville}, \bibinfo{author}{B.~Escaig},
  \bibinfo{author}{J.~Martin},
\newblock \bibinfo{title}{A study of cross-slip activation parameters in pure
  copper},
\newblock \bibinfo{journal}{Acta Metallurgica} \bibinfo{volume}{36}
  (\bibinfo{year}{1988}) \bibinfo{pages}{1989--2002}.
\bibitem[{Rasmussen et~al.(1997)Rasmussen, Jacobsen, Leffers, and
  Pedersen}]{rasmussen1997}
\bibinfo{author}{T.~Rasmussen}, \bibinfo{author}{K.~W. Jacobsen},
  \bibinfo{author}{T.~Leffers}, \bibinfo{author}{O.~B. Pedersen},
\newblock \bibinfo{title}{Simulations of the atomic structure, energetics, and
  cross slip of screw dislocations in copper},
\newblock \bibinfo{journal}{Phys. Rev. B} \bibinfo{volume}{56}
  (\bibinfo{year}{1997}) \bibinfo{pages}{2977--2990}.
\bibitem[{Nabarro(1947)}]{nabarro1947}
\bibinfo{author}{F.~R.~N. Nabarro},
\newblock \bibinfo{title}{Dislocations in a simple cubic lattice},
\newblock \bibinfo{journal}{Proceedings of the Physical Society}
  \bibinfo{volume}{59} (\bibinfo{year}{1947}) \bibinfo{pages}{256--272}.
\bibitem[{Vitek(1968)}]{vitek1968}
\bibinfo{author}{V.~Vitek},
\newblock \bibinfo{title}{Intrinsic stacking faults in body-centred cubic
  crystals},
\newblock \bibinfo{journal}{The Philosophical Magazine: A Journal of
  Theoretical Experimental and Applied Physics} \bibinfo{volume}{18}
  (\bibinfo{year}{1968}) \bibinfo{pages}{773--786}.
\bibitem[{Bulatov and Kaxiras(1997)}]{bulatov1997}
\bibinfo{author}{V.~V. Bulatov}, \bibinfo{author}{E.~Kaxiras},
\newblock \bibinfo{title}{Semidiscrete variational peierls framework for
  dislocation core properties},
\newblock \bibinfo{journal}{Phys. Rev. Lett.} \bibinfo{volume}{78}
  (\bibinfo{year}{1997}) \bibinfo{pages}{4221--4224}.
\bibitem[{Schoeck(2005)}]{schoeck20057}
\bibinfo{author}{G.~Schoeck},
\newblock \bibinfo{title}{The peierls model: Progress and limitations},
\newblock \bibinfo{journal}{Materials Science and Engineering: A}
  \bibinfo{volume}{400-401} (\bibinfo{year}{2005}) \bibinfo{pages}{7--17}.
\bibitem[{Stobbs and Sworn(1971)}]{stobbs1971}
\bibinfo{author}{W.~M. Stobbs}, \bibinfo{author}{C.~H. Sworn},
\newblock \bibinfo{title}{The weak beam technique as applied to the
  determination of the stacking-fault energy of copper},
\newblock \bibinfo{journal}{The Philosophical Magazine: A Journal of
  Theoretical Experimental and Applied Physics} \bibinfo{volume}{24}
  (\bibinfo{year}{1971}) \bibinfo{pages}{1365--1381}.
\bibitem[{Weiler et~al.(1995)Weiler, Sigle, and Seeger}]{weiler1995high}
\bibinfo{author}{B.~Weiler}, \bibinfo{author}{W.~Sigle},
  \bibinfo{author}{A.~Seeger},
\newblock \bibinfo{title}{High-resolution electron-microscopy study of
  60°-dislocations in cu},
\newblock \bibinfo{journal}{Physica Status Solidi (A)} \bibinfo{volume}{150}
  (\bibinfo{year}{1995}) \bibinfo{pages}{221--225}.
\bibitem[{Rodney et~al.(2017)Rodney, Ventelon, Clouet, Pizzagalli, and
  Willaime}]{rodney2017}
\bibinfo{author}{D.~Rodney}, \bibinfo{author}{L.~Ventelon},
  \bibinfo{author}{E.~Clouet}, \bibinfo{author}{L.~Pizzagalli},
  \bibinfo{author}{F.~Willaime},
\newblock \bibinfo{title}{Ab initio modeling of dislocation core properties in
  metals and semiconductors},
\newblock \bibinfo{journal}{Acta Materialia} \bibinfo{volume}{124}
  (\bibinfo{year}{2017}) \bibinfo{pages}{633--659}.
\bibitem[{Gr{\"o}ger and Vitek(2009)}]{groger2009}
\bibinfo{author}{R.~Gr{\"o}ger}, \bibinfo{author}{V.~Vitek},
\newblock \bibinfo{title}{Directional versus central-force bonding in studies
  of the structure and glide of 1/2⟨111⟩ screw dislocations in bcc
  transition metals},
\newblock \bibinfo{journal}{Philosophical Magazine} \bibinfo{volume}{89}
  (\bibinfo{year}{2009}) \bibinfo{pages}{3163--3178}.
\bibitem[{Chiesa et~al.(2009)Chiesa, Gilbert, Dudarev, Derlet, and
  Swygenhoven}]{chiesa2009}
\bibinfo{author}{S.~Chiesa}, \bibinfo{author}{M.~Gilbert},
  \bibinfo{author}{S.~Dudarev}, \bibinfo{author}{P.~Derlet},
  \bibinfo{author}{H.~V. Swygenhoven},
\newblock \bibinfo{title}{The non-degenerate core structure of a ½⟨111⟩
  screw dislocation in bcc transition metals modelled using finnis–sinclair
  potentials: The necessary and sufficient conditions},
\newblock \bibinfo{journal}{Philosophical Magazine} \bibinfo{volume}{89}
  (\bibinfo{year}{2009}) \bibinfo{pages}{3235--3243}.
\bibitem[{Behler and Parrinello(2007)}]{behler2007ge}
\bibinfo{author}{J.~Behler}, \bibinfo{author}{M.~Parrinello},
\newblock \bibinfo{title}{Generalized neural-network representation of
  high-dimensional potential-energy surfaces},
\newblock \bibinfo{journal}{Phys. Rev. Lett.} \bibinfo{volume}{98}
  (\bibinfo{year}{2007}) \bibinfo{pages}{146401}.
\bibitem[{Bart\'ok et~al.(2010)Bart\'ok, Payne, Kondor, and
  Cs\'anyi}]{bartok2010}
\bibinfo{author}{A.~P. Bart\'ok}, \bibinfo{author}{M.~C. Payne},
  \bibinfo{author}{R.~Kondor}, \bibinfo{author}{G.~Cs\'anyi},
\newblock \bibinfo{title}{Gaussian approximation potentials: The accuracy of
  quantum mechanics, without the electrons},
\newblock \bibinfo{journal}{Phys. Rev. Lett.} \bibinfo{volume}{104}
  (\bibinfo{year}{2010}) \bibinfo{pages}{136403}.
\bibitem[{Zhao et~al.(2020)Zhao, Fan, Su, Song, Wang, and Qiao}]{zhao2020snap}
\bibinfo{author}{Y.~Zhao}, \bibinfo{author}{J.~Fan}, \bibinfo{author}{L.~Su},
  \bibinfo{author}{T.~Song}, \bibinfo{author}{S.~Wang},
  \bibinfo{author}{C.~Qiao},
\newblock \bibinfo{title}{Snap: A communication efficient distributed machine
  learning framework for edge computing},
\newblock in: \bibinfo{booktitle}{2020 IEEE 40th International Conference on
  Distributed Computing Systems (ICDCS)}, \bibinfo{year}{2020}, pp.
  \bibinfo{pages}{584--594}. \DOIprefix\doi{10.1109/ICDCS47774.2020.00072}.
\bibitem[{Sch{\"u}tt et~al.(2018)Sch{\"u}tt, Sauceda, Kindermans, Tkatchenko,
  and M{\"u}ller}]{schutt2018schnet}
\bibinfo{author}{K.~T. Sch{\"u}tt}, \bibinfo{author}{H.~E. Sauceda},
  \bibinfo{author}{P.-J. Kindermans}, \bibinfo{author}{A.~Tkatchenko},
  \bibinfo{author}{K.-R. M{\"u}ller},
\newblock \bibinfo{title}{Schnet--a deep learning architecture for molecules
  and materials},
\newblock \bibinfo{journal}{The Journal of Chemical Physics}
  \bibinfo{volume}{148} (\bibinfo{year}{2018}) \bibinfo{pages}{241722}.
\bibitem[{Wang et~al.(2018)Wang, Zhang, Han, and E}]{wang2018dpkit}
\bibinfo{author}{H.~Wang}, \bibinfo{author}{L.~Zhang},
  \bibinfo{author}{J.~Han}, \bibinfo{author}{W.~E},
\newblock \bibinfo{title}{{DeePMD-kit: A deep learning package for many-body
  potential energy representation and molecular dynamics}},
\newblock \bibinfo{journal}{Comput. Phys. Commun.} \bibinfo{volume}{228}
  (\bibinfo{year}{2018}) \bibinfo{pages}{178--184}.
\bibitem[{Zhang et~al.(2018)Zhang, Han, Wang, Car, and E}]{zhang2018dp}
\bibinfo{author}{L.~Zhang}, \bibinfo{author}{J.~Han},
  \bibinfo{author}{H.~Wang}, \bibinfo{author}{R.~Car}, \bibinfo{author}{W.~E},
\newblock \bibinfo{title}{Deep potential molecular dynamics: A scalable model
  with the accuracy of quantum mechanics},
\newblock \bibinfo{journal}{Phys. Rev. Lett.} \bibinfo{volume}{120}
  (\bibinfo{year}{2018}) \bibinfo{pages}{143001}.
\bibitem[{Zhang et~al.(2021)Zhang, Wang, Car, and E}]{zhang2021pd}
\bibinfo{author}{L.~Zhang}, \bibinfo{author}{H.~Wang},
  \bibinfo{author}{R.~Car}, \bibinfo{author}{W.~E},
\newblock \bibinfo{title}{Phase diagram of a deep potential water model},
\newblock \bibinfo{journal}{Phys. Rev. Lett.} \bibinfo{volume}{126}
  (\bibinfo{year}{2021}) \bibinfo{pages}{236001}.
\bibitem[{Zhang et~al.(2018)Zhang, Han, Wang, Saidi, Car, and E}]{zhang2018end}
\bibinfo{author}{L.~Zhang}, \bibinfo{author}{J.~Han},
  \bibinfo{author}{H.~Wang}, \bibinfo{author}{W.~A. Saidi},
  \bibinfo{author}{R.~Car}, \bibinfo{author}{W.~E},
\newblock \bibinfo{title}{End-to-end symmetry preserving inter-atomic potential
  energy model for finite and extended systems},
\newblock in: \bibinfo{booktitle}{Proceedings of the 32nd International
  Conference on Neural Information Processing Systems}, \bibinfo{year}{2018},
  pp. \bibinfo{pages}{4441--4451}. \DOIprefix\doi{10.5555/3327345.3327356}.
\bibitem[{Fu et~al.(2021)Fu, Sun, Zhang, Wang, and Xu}]{fu2021deep}
\bibinfo{author}{B.~Fu}, \bibinfo{author}{Y.~Sun}, \bibinfo{author}{L.~Zhang},
  \bibinfo{author}{H.~Wang}, \bibinfo{author}{B.~Xu}, \bibinfo{title}{Deep
  learning inter-atomic potential for thermal and phonon behaviour of silicon
  carbide with quantum accuracy}, \bibinfo{year}{2021}.
  \href{http://arxiv.org/abs/2110.10843}{\tt arXiv:2110.10843}.
\bibitem[{He et~al.(2022)He, Wu, Zhang, Wang, Fu, Liu, and
  Zhong}]{he2022structural}
\bibinfo{author}{R.~He}, \bibinfo{author}{H.~Wu}, \bibinfo{author}{L.~Zhang},
  \bibinfo{author}{X.~Wang}, \bibinfo{author}{F.~Fu}, \bibinfo{author}{S.~Liu},
  \bibinfo{author}{Z.~Zhong},
\newblock \bibinfo{title}{Structural phase transitions in
  $\mathrm{SrTi}{\mathrm{o}}_{3}$ from deep potential molecular dynamics},
\newblock \bibinfo{journal}{Phys. Rev. B} \bibinfo{volume}{105}
  (\bibinfo{year}{2022}) \bibinfo{pages}{064104}.
\bibitem[{Zhang et~al.(2020)Zhang, Wang, Chen, Zeng, Zhang, Wang, and
  E}]{zhang2020}
\bibinfo{author}{Y.~Zhang}, \bibinfo{author}{H.~Wang},
  \bibinfo{author}{W.~Chen}, \bibinfo{author}{J.~Zeng},
  \bibinfo{author}{L.~Zhang}, \bibinfo{author}{H.~Wang},
  \bibinfo{author}{W.~E},
\newblock \bibinfo{title}{Dp-gen: A concurrent learning platform for the
  generation of reliable deep learning based potential energy models},
\newblock \bibinfo{journal}{Computer Physics Communications}
  \bibinfo{volume}{253} (\bibinfo{year}{2020}) \bibinfo{pages}{107206}.
\bibitem[{Jiang et~al.(2021)Jiang, Zhang, Zhang, and Wang}]{jiang2021}
\bibinfo{author}{W.~Jiang}, \bibinfo{author}{Y.~Zhang},
  \bibinfo{author}{L.~Zhang}, \bibinfo{author}{H.~Wang},
\newblock \bibinfo{title}{Accurate deep potential model for the al-cu-mg alloy
  in the full concentration space},
\newblock \bibinfo{journal}{Chinese Physics B} \bibinfo{volume}{30}
  (\bibinfo{year}{2021}) \bibinfo{pages}{050706}.
\bibitem[{Wang et~al.(2015)Wang, Zhang, Bai, and Yao}]{wang2015}
\bibinfo{author}{S.~Wang}, \bibinfo{author}{S.~Zhang},
  \bibinfo{author}{J.~Bai}, \bibinfo{author}{Y.~Yao},
\newblock \bibinfo{title}{Shape change and peierls barrier of dislocation},
\newblock \bibinfo{journal}{Journal of Applied Physics} \bibinfo{volume}{118}
  (\bibinfo{year}{2015}) \bibinfo{pages}{244903}.
\bibitem[{Wang et~al.(2016)Wang, Huang, and Wang}]{wang2016}
\bibinfo{author}{S.~Wang}, \bibinfo{author}{L.~Huang},
  \bibinfo{author}{R.~Wang},
\newblock \bibinfo{title}{The 90° partial dislocation in semiconductor
  silicon: An investigation from the lattice p–n theory and the first
  principle calculation},
\newblock \bibinfo{journal}{Acta Materialia} \bibinfo{volume}{109}
  (\bibinfo{year}{2016}) \bibinfo{pages}{187--201}.
\bibitem[{Xiang et~al.(2020)Xiang, Wang, and Wang}]{xiang2020}
\bibinfo{author}{H.~Xiang}, \bibinfo{author}{R.~Wang},
  \bibinfo{author}{S.~Wang},
\newblock \bibinfo{title}{Core structure and thermal transformation of the
  1/2$\langle 110 \rangle$$\{$111$\}$ screw dislocation in aluminum},
\newblock \bibinfo{journal}{Journal of Applied Physics} \bibinfo{volume}{127}
  (\bibinfo{year}{2020}) \bibinfo{pages}{125106}.
\bibitem[{LeSar(2014)}]{lesar2014}
\bibinfo{author}{R.~LeSar},
\newblock \bibinfo{title}{Simulations of dislocation structure and response},
\newblock \bibinfo{journal}{Annual Review of Condensed Matter Physics}
  \bibinfo{volume}{5} (\bibinfo{year}{2014}) \bibinfo{pages}{375--407}.
\bibitem[{Beyerlein and Hunter(2016)}]{beyerlein2016under}
\bibinfo{author}{I.~Beyerlein}, \bibinfo{author}{A.~Hunter},
\newblock \bibinfo{title}{Understanding dislocation mechanics at the mesoscale
  using phase field dislocation dynamics},
\newblock \bibinfo{journal}{Philosophical Transactions of the Royal Society A:
  Mathematical, Physical and Engineering Sciences} \bibinfo{volume}{374}
  (\bibinfo{year}{2016}) \bibinfo{pages}{20150166}.
\bibitem[{Bertin et~al.(2020)Bertin, Sills, and Cai}]{bertin2020}
\bibinfo{author}{N.~Bertin}, \bibinfo{author}{R.~B. Sills},
  \bibinfo{author}{W.~Cai},
\newblock \bibinfo{title}{Frontiers in the simulation of dislocations},
\newblock \bibinfo{journal}{Annual Review of Materials Research}
  \bibinfo{volume}{50} (\bibinfo{year}{2020}) \bibinfo{pages}{437--464}.
\bibitem[{Deng et~al.(2019)Deng, Hu, and Wang}]{deng2019}
\bibinfo{author}{F.~Deng}, \bibinfo{author}{X.~Hu}, \bibinfo{author}{S.~Wang},
\newblock \bibinfo{title}{Dislocation neutralizing in a self-organized array of
  dislocation and anti-dislocation},
\newblock \bibinfo{journal}{Chinese Physics B} \bibinfo{volume}{28}
  (\bibinfo{year}{2019}) \bibinfo{pages}{116103}.
\bibitem[{Kresse and Furthmüller(1996)}]{kresse199615}
\bibinfo{author}{G.~Kresse}, \bibinfo{author}{J.~Furthmüller},
\newblock \bibinfo{title}{Efficiency of $ab$-$initio$ total energy calculations
  for metals and semiconductors using a plane-wave basis set},
\newblock \bibinfo{journal}{Computational Materials Science}
  \bibinfo{volume}{6} (\bibinfo{year}{1996}) \bibinfo{pages}{15 -- 50}.
\bibitem[{Kresse and Furthm\"uller(1996)}]{kresse199610}
\bibinfo{author}{G.~Kresse}, \bibinfo{author}{J.~Furthm\"uller},
\newblock \bibinfo{title}{Efficient iterative schemes for $ab$ $initio$
  total-energy calculations using a plane-wave basis set},
\newblock \bibinfo{journal}{Phys. Rev. B} \bibinfo{volume}{54}
  (\bibinfo{year}{1996}) \bibinfo{pages}{11169--11186}.
\bibitem[{Csonka et~al.(2009)Csonka, Perdew, Ruzsinszky, Philipsen, Leb\`egue,
  Paier, Vydrov, and \'Angy\'an}]{PBEsol2009}
\bibinfo{author}{G.~I. Csonka}, \bibinfo{author}{J.~P. Perdew},
  \bibinfo{author}{A.~Ruzsinszky}, \bibinfo{author}{P.~H.~T. Philipsen},
  \bibinfo{author}{S.~Leb\`egue}, \bibinfo{author}{J.~Paier},
  \bibinfo{author}{O.~A. Vydrov}, \bibinfo{author}{J.~G. \'Angy\'an},
\newblock \bibinfo{title}{Assessing the performance of recent density
  functionals for bulk solids},
\newblock \bibinfo{journal}{Phys. Rev. B} \bibinfo{volume}{79}
  (\bibinfo{year}{2009}) \bibinfo{pages}{155107}.
\bibitem[{Monkhorst and Pack(1976)}]{monkhorst1976}
\bibinfo{author}{H.~J. Monkhorst}, \bibinfo{author}{J.~D. Pack},
\newblock \bibinfo{title}{Special points for brillouin-zone integrations},
\newblock \bibinfo{journal}{Phys. Rev. B} \bibinfo{volume}{13}
  (\bibinfo{year}{1976}) \bibinfo{pages}{5188--5192}.
\bibitem[{Plimpton(1995)}]{plimpton19951}
\bibinfo{author}{S.~Plimpton},
\newblock \bibinfo{title}{Fast parallel algorithms for short-range molecular
  dynamics},
\newblock \bibinfo{journal}{Journal of Computational Physics}
  \bibinfo{volume}{117} (\bibinfo{year}{1995}) \bibinfo{pages}{1--19}.
\bibitem[{Mishin et~al.(2001)Mishin, Mehl, Papaconstantopoulos, Voter, and
  Kress}]{mishin2001}
\bibinfo{author}{Y.~Mishin}, \bibinfo{author}{M.~J. Mehl},
  \bibinfo{author}{D.~A. Papaconstantopoulos}, \bibinfo{author}{A.~F. Voter},
  \bibinfo{author}{J.~D. Kress},
\newblock \bibinfo{title}{Structural stability and lattice defects in copper:
  Ab initio, tight-binding, and embedded-atom calculations},
\newblock \bibinfo{journal}{Phys. Rev. B} \bibinfo{volume}{63}
  (\bibinfo{year}{2001}) \bibinfo{pages}{224106}.
\bibitem[{Duesbery and Vitek(1998)}]{DUESBERY19981481}
\bibinfo{author}{M.~Duesbery}, \bibinfo{author}{V.~Vitek},
\newblock \bibinfo{title}{Plastic anisotropy in b.c.c. transition metals},
\newblock \bibinfo{journal}{Acta Materialia} \bibinfo{volume}{46}
  (\bibinfo{year}{1998}) \bibinfo{pages}{1481--1492}.
\bibitem[{Wang(2008)}]{wang2008dis}
\bibinfo{author}{S.~F. Wang},
\newblock \bibinfo{title}{A unified dislocation equation from lattice statics},
\newblock \bibinfo{journal}{Journal of Physics A: Mathematical and Theoretical}
  \bibinfo{volume}{42} (\bibinfo{year}{2008}) \bibinfo{pages}{025208}.
\bibitem[{Vorontsov et~al.(2012)Vorontsov, Voskoboinikov, and
  Rae}]{vorontsov2012}
\bibinfo{author}{V.~Vorontsov}, \bibinfo{author}{R.~Voskoboinikov},
  \bibinfo{author}{C.~Rae},
\newblock \bibinfo{title}{Shearing of $\gamma^{\prime}$ precipitates in ni-base
  superalloys: a phase field study incorporating the effective
  $\gamma$-surface},
\newblock \bibinfo{journal}{Philosophical Magazine} \bibinfo{volume}{92}
  (\bibinfo{year}{2012}) \bibinfo{pages}{608--634}.
\bibitem[{Karlin et~al.(2000)Karlin, Maz'ya, Movchan, Willis, and
  Bullough}]{karlin2000numerical}
\bibinfo{author}{V.~Karlin}, \bibinfo{author}{V.~Maz'ya},
  \bibinfo{author}{A.~Movchan}, \bibinfo{author}{J.~Willis},
  \bibinfo{author}{R.~Bullough},
\newblock \bibinfo{title}{Numerical solution of nonlinear hypersingular
  integral equations of the peierls type in dislocation theory},
\newblock \bibinfo{journal}{SIAM Journal on Applied Mathematics}
  \bibinfo{volume}{60} (\bibinfo{year}{2000}) \bibinfo{pages}{664--678}.
\bibitem[{Liu et~al.(2017{\natexlab{a}})Liu, Cheng, Wang, Chen, and
  Shen}]{liu2017156}
\bibinfo{author}{G.~Liu}, \bibinfo{author}{X.~Cheng},
  \bibinfo{author}{J.~Wang}, \bibinfo{author}{K.~Chen},
  \bibinfo{author}{Y.~Shen},
\newblock \bibinfo{title}{Quasi-periodic variation of peierls stress of
  dislocations in face-centered-cubic metals},
\newblock \bibinfo{journal}{International Journal of Plasticity}
  \bibinfo{volume}{90} (\bibinfo{year}{2017}{\natexlab{a}})
  \bibinfo{pages}{156--166}.
\bibitem[{Liu et~al.(2017{\natexlab{b}})Liu, Cheng, Wang, Chen, and
  Shen}]{liu201769}
\bibinfo{author}{G.~Liu}, \bibinfo{author}{X.~Cheng},
  \bibinfo{author}{J.~Wang}, \bibinfo{author}{K.~Chen},
  \bibinfo{author}{Y.~Shen},
\newblock \bibinfo{title}{Improvement of nonlocal peierls-nabarro models},
\newblock \bibinfo{journal}{Computational Materials Science}
  \bibinfo{volume}{131} (\bibinfo{year}{2017}{\natexlab{b}})
  \bibinfo{pages}{69--77}.
\bibitem[{Dutta et~al.(2008)Dutta, Bhattacharya, Barat, Mukherjee, Gayathri,
  and Das}]{dutta2008}
\bibinfo{author}{A.~Dutta}, \bibinfo{author}{M.~Bhattacharya},
  \bibinfo{author}{P.~Barat}, \bibinfo{author}{P.~Mukherjee},
  \bibinfo{author}{N.~Gayathri}, \bibinfo{author}{G.~C. Das},
\newblock \bibinfo{title}{Lattice resistance to dislocation motion at the
  nanoscale},
\newblock \bibinfo{journal}{Phys. Rev. Lett.} \bibinfo{volume}{101}
  (\bibinfo{year}{2008}) \bibinfo{pages}{115506}.
\bibitem[{Cheng et~al.(2012)Cheng, Shen, Zhang, and Liu}]{cheng2012}
\bibinfo{author}{X.~Cheng}, \bibinfo{author}{Y.~Shen},
  \bibinfo{author}{L.~Zhang}, \bibinfo{author}{X.~Liu},
\newblock \bibinfo{title}{Surface effect on the screw dislocation mobility over
  the peierls barrier},
\newblock \bibinfo{journal}{Philosophical Magazine Letters}
  \bibinfo{volume}{92} (\bibinfo{year}{2012}) \bibinfo{pages}{270--277}.
\bibitem[{Bai and Wang(2016)}]{bai2016}
\bibinfo{author}{J.~Bai}, \bibinfo{author}{S.~Wang},
\newblock \bibinfo{title}{Screw dislocation equations in a thin film and
  surface effects},
\newblock \bibinfo{journal}{International Journal of Plasticity}
  \bibinfo{volume}{87} (\bibinfo{year}{2016}) \bibinfo{pages}{181--203}.
\bibitem[{Bullough and Newman(1970)}]{bullough1970}
\bibinfo{author}{R.~Bullough}, \bibinfo{author}{R.~C. Newman},
\newblock \bibinfo{title}{The kinetics of migration of point defects to
  dislocations},
\newblock \bibinfo{journal}{Reports on Progress in Physics}
  \bibinfo{volume}{33} (\bibinfo{year}{1970}) \bibinfo{pages}{101--148}.
\bibitem[{Clouet(2006)}]{clouet20063543}
\bibinfo{author}{E.~Clouet},
\newblock \bibinfo{title}{The vacancy–edge dislocation interaction in fcc
  metals: A comparison between atomic simulations and elasticity theory},
\newblock \bibinfo{journal}{Acta Materialia} \bibinfo{volume}{54}
  (\bibinfo{year}{2006}) \bibinfo{pages}{3543--3552}. \bibinfo{note}{Selected
  Papers from the Meeting “Micromechanics and Microstructure Evolution:
  Modeling, Simulation and Experiments” held in Madrid/Spain, 11–16
  September 2005}.

\end{thebibliography}

\end{document}